\documentstyle[preprint,aps]{revtex}
\begin {document}
\title{\bf Photodisintegration of Three-Body Nuclei with
Realistic 2N and 3N Forces}

\author{Victor D. Efros$^{1}$, Winfried Leidemann$^{2}$,
 Giuseppina Orlandini$^{2}$ and Edward L. Tomusiak$^{3}$}
\address{
$^{1}$Russian Research Centre "Kurchatov Institute",Kurchatov Square, 1, 123182 Moscow, Russia\\
$^{2}$Dipartimento di Fisica, Universit\`a di Trento, and\\
Istituto Nazionale di Fisica Nucleare, Gruppo collegato di Trento,
 I-38050 Povo, Italy\\
$^{3}$Department of Physics and Engineering Physics\\
University of Saskatchewan, Saskatoon, Canada S7N 0W0
}

\date{\today}
\maketitle

\begin{abstract}
Total photonuclear absorption cross sections of $^3$H
and $^3$He are studied using realistic NN and NNN forces.
Final state interactions are fully included.
Two NN potential models, the AV14 and the r-space Bonn-A 
potentials, are considered.  For the NNN forces the Urbana-VIII and
Tucson-Melbourne models are employed.  
We find the cross section
to be sensitive to nuclear dynamics. Of particular
interest in this work is the effect which NNN forces
have on the cross section. The
addition of NNN forces not only lowers the peak height but
increases the cross section beyond 70 MeV by roughly 15\%.
Cross sections are computed using the Lorentz
integral transform method.
\end{abstract}
\bigskip
\pacs{PACS numbers: 25.20.Dc, 21.30-x, 21.45.+v, 27.10.+h\\
Keywords: photoabsorption; three-body forces; few-body physics}

\vfill\eject

The process of nuclear photoabsorption is known to be strongly
dependent on exchange currents and hence on the underlying
nuclear dynamics.  One therefore expects the photoabsorption
cross section to be sensitive to NN and NNN nuclear forces. 
It is the purpose of this paper to explore
this sensitivity by computing the total photoabsorption cross sections
of the trinucleons by using realistic NN and NNN potentials. The only 
other calculations of trinucleon photodisintegraton covering a larger range
of energies (0 - 40 MeV) and employing 
realistic NN forces are, to our knowledge, those of Sandhas {\it et al}
\cite{sandhas}.  However their calculations are for the exclusive
process of two-body breakup while we compute the inclusive
 (two-body + three-body breakup) cross sections over a larger
energy range.  It is because we are looking for effects of NN and
NNN forces that energies in the tail region, E$_\gamma >$70 MeV, are 
included here.

We use the Argonne AV14
potential \cite{av14} and the Bonn-A r-space potential \cite{bonnb}. 
As NNN potentials we include the Urbana-VIII (UrbVIII)
\cite{urb8} and Tucson-Melbourne (TM) \cite{TM} NNN models.
The total photoabsorption cross section for energies below
pion threshold is computed by using the unretarded dipole operator 
\begin{eqnarray}
\vec D\,=\,\sum_{i=1}^Z\,(\vec r_i\,-\,\vec R_{cm})\ .
\end{eqnarray}
This form of the operator already includes the effects of exchange
currents due to Siegert's theorem.  
It is known that as far as the total cross section is concerned
corrections to the unretarded dipole operator are small 
in the energy range studied here.
We consider this approximation  adequate for investigating the
effects of NN and NNN forces to the process. 
The total photoabsorption cross section
is then given in terms of the dipole response function {\it R} by
\begin{eqnarray}
\sigma_{tot}\,=\,4\pi^2(e^2/\hbar c)E_\gamma\,R(E_\gamma)
\end{eqnarray}
where
\begin{eqnarray}
R(\omega)\,=\,<\Psi_0|D_z \delta(H-E_0-\omega) D_z|\Psi_0>\  .
\end{eqnarray}
To calculate $R(\omega)$ we use the Lorentz integral transform
method introduced in \cite{ELO1}. This technique eliminates the
need to compute final-state continuum wave functions.
The transform, $\Phi(\sigma_R,\sigma_I)$,  of $R(\omega)$ 
is defined by
\begin{eqnarray}
\Phi(\sigma_R,\sigma_I)\ =\ \int_{\omega_{min}}^\infty\ d\omega\ 
{R(\omega)\over (\omega - \sigma_R)^2 + \sigma_I^2}\ \  .
\end{eqnarray}
It can be expressed as the norm of the square integrable function $\tilde\Psi$
\begin{eqnarray}
\Phi(\sigma_R,\sigma_I)\ =\ <\tilde\Psi|\tilde\Psi>
\end{eqnarray}
where 
\begin{eqnarray}
(\, H-E_0-\sigma_R + i\sigma_I\,)|\tilde\Psi>\ =\ D_z |\Psi_0>.
\end{eqnarray}
Finally $R(\omega)$ is obtained by inverting
the transform (4).  The inversion procedure is elaborated upon
in \cite{ELO1,ELO2,ELO3}.  This method was applied to the study
of few-nucleon responses in \cite{ELO2,mkog,ELO4,ELO6,ELO5}. 
In \cite{mkog} Faddeev techniques were applied to solve
equations of the form of Eq.(6)
whereas \cite{ELO2,ELO4,ELO6,ELO5} used expansions in a set of
correlated hyperspherical harmonics (CHH) for this purpose.
The latter technique is also adopted in the present work.
We expand the functions, $\Psi_0$ and $\tilde\Psi$,
in a set of correlated hyperspherical harmonics (CHH)
according to
\begin{equation}
\Psi\ =\ \tilde\omega\,\sum\,c_i\,\phi_i
\end{equation}
where $\tilde\omega$ is a correlation operator and
$\phi_i$ are a totally antisymmetric basis set constructed from
a spatial part $\chi_{i,\mu}$ and a spin-isospin part $\theta_{\mu}$ 
\begin{equation}
\phi_i\ =\ \sum_{\mu}\,\chi_{i,\mu} \theta_{\mu}\ .
\end{equation}

Whereas in \cite{ELO2,ELO4,ELO6}, where simpler potential models were
employed, it was sufficient to take $\tilde\omega$
in the state independent form
\begin{equation}
\tilde\omega\ =\ \prod_{i<j}\,f(r_{ij})\ .
\end{equation}
we now use a state dependent correlation operator of the form
\begin{equation}
\tilde\omega\ =\ {\cal S}\prod_{i<j}\,\sum_{S,T}\,f_{ST}(r_{ij})\,\tilde P_{ST}(ij)
\end{equation}
where $P_{ST}(ij)$ are projection operators onto nucleon pairs (ij) with spin $S$
and isospin $T$ and where ${\cal S}$ is a particle symmetrization operator.
This form of correlation operator is easily incorporated into the calculations
by first constructing the set of coefficients $<\theta_{\mu^\prime}|\tilde\omega
|\theta_{\mu}>$ and then writing
\begin{equation}
\Psi\ =\ \sum_i\,\tilde\chi_{i,\mu^\prime}\,\theta_{\mu^\prime}
\end{equation}
where
\begin{equation}
\tilde\chi_{i,\mu^\prime}\ =\ \sum_{\mu} <\theta_{\mu^\prime}|\tilde\omega |\theta_{\mu}> \,\chi_{i,\mu}\ .
\end{equation}
The correlation functions $f_{ST}(r)$ are chosen as follows. 
For $r<r_0$ , the healing distance, $f_{ST}(r)$ is chosen to be
the zero energy pair wave function in
the corresponding $ST$ state.  Healing is insured by imposing the conditions
$f_{ST}(r)$=1 for $r>r_0$ and  $f_{ST}^\prime (r_0)$=0.
The $ST$= 13 and 31 cases are determined from the $^1$S$_0$ and
$^3$S$_1$ partial waves of the NN potential.
However for the $ST$=11 and 33 cases the  $^1$P$_1$ and $^3$P$_1$
potentials are not sufficiently attractive to obtain a healing distance.
Therefore we introduce an additional intermediate range central
interaction such that a healing distance of 10 fm is obtained.  
Further details can be found in \cite{ELO5}.
 
Table 1 gives our results for the ground state properties of $^3$H using
the various potential models described above.  We point out that
our results for both the ground state and the Lorentz integral
transform arise from fully converged expansions in the CHH basis.
The inversion leads to stable results for $R(\omega)$.
In order to obtain the
correct binding energy with the TM NNN potential we have 
adjusted the cut-off mass $\Lambda$ in the monopole form factor.  This requires
$\Lambda$= 4.67$\mu$ and $\Lambda$=4.07$\mu$ for use with
the AV14 and Bonn-A (r-space) cases respectively.  Previously published
ground state properties are available \cite{chen,ariaga} for the AV14
and AV14+UrbVIII potentials. The corresponding results in Table 1 are
in agreement with these.
Since the method of the Lorentz integral transform allows one to 
calculate the cross section over the entire energy range, sum rules
can be used to give additional checks.  
An example is provided by  explicit evaluation
of the inverse energy weighted sum rule
$\sigma_{-1}=\int_{E_{th}}^\infty\,E_\gamma^{-1} \sigma_{tot}(E_\gamma)
dE_\gamma$ which is related to the triton point proton radius
through
$$
\sigma_{-1}(^3{\rm H})\ =\ {4\pi^2\over 3}{e^2\over\hbar c}<r_p^2(^3{\rm H})>\ .
$$
This sum rule is calculated in two ways: (i) by direct integration
of the response giving $\sigma_{-1}^{int}$, and (ii) by 
use of the ground state wave function giving $\sigma_{-1}^{gs}$.
As seen in Table 1 agreement is obtained to better than 0.5\%.  This
level of agreement helps confirm that the lower
energy regions, and especially the peak heights, are correct. 
A check on the higher energy regions is provided by evaluating
the TRK sum rule, again by direct integration and by the ground
state expectation value of the appropriate double commutator.
This was done for the AV14 model, yielding 70.7 MeV-mb
by direct integration and 71.6 MeV-mb by evaluating the double
commutator. The direct integration was carried out to 2000 MeV.
It should be noted that integrating up to 300 MeV excitation
only exhausts about 90\% of the sum rule.  This is in contrast
to the integration of the TRK sum rule produced by a soft-core
potential such as TRSB potential \cite{TRSB} where almost all the strength
is exhausted at 300 MeV excitation \cite{ELO2}.

The value of the cross section at the peak depends
on ground state details, especially the radius.  This
is evident from the remarks above on the inverse energy
weighted sum rule.  Fig.1, for example, shows 
results obtained previously \cite{ELO4} using the
Malfliet-Tjon (MT) \cite{mt} and
Trento (TN) \cite{ELO2} potentials together with
new results obtained here for the AV14 and AV14+UrbVIII potentials . 
Ground state point proton radii computed from these potentials
are 1.61, 1.59, 1.66, and 1.60 fm
respectively. The similar peak heights for these rather
different potentials is due largely to the similarity
of these radii. However, we have observed that in all cases considered
here the combination of realistic NN forces with NNN forces leads
to roughly a 10 \% decrease in the peak height.
Moreover, even though the four potentials 
give similar radii and binding energies the 
AV14 and AV14+UrbVIII potentials produce cross sections
with significantly larger high energy tails than the semi-realistic MT
and TN models. Part of the difference must be attributable to
the tensor force contained in the realistic potential
models.

  Figs.2(a) and 2(b) show the peak and tail regions of the T=1/2
contribution to the absorption cross section as computed from the
five potential models of Table 1. A study of these figures
reveals that the addition of three-body forces (3BF) lowers the peak height
and increases the cross section in the tail region by 10 - 20\%.
The lowering of the peak can be
understood since the addition of 3BF increases the binding energy
and decreases the radius in all cases. 
Despite the increase of the cross section in the tail region due
to 3BF, no clear separation between the cross section with and
without 3BF is seen in this channel.
 We note the importance of binding
in this T=1/2 channel by observing in the tail region the rather disparate
curves for the pure NN cases of AV14 and Bonn-A.  Recall from Table 1
that these NN potentials correspond to $^3$H binding energies of
7.69 MeV and 8.15 MeV respectively.  In addition the three curves
corresponding to the addition of NNN forces are rather close and
difficult to distinguish, reinforcing the argument that effects here
are mainly due to binding.

  Curves corresponding to Fig.2 but for the T=3/2 channel are 
shown in Figs.3(a) and 3(b).  The behaviour of the
cross section in the peak region is similar to the T=1/2 case
although here the peak height is less sensitive to the interaction
model.
Fig.3(b) shows that the T=3/2 channel
clearly separates all models containing a 3BF from those with
purely NN forces.  This is a 5 - 10 \% effect in the energy
range 70-110 MeV.  However contrary to
the T=1/2 case, binding effects here do not play an important role
as evidenced by the nearly overlapping NN curves in the tail region.
As a result the separation between the curves in Fig.3(b)
is dominantly an effect due to NNN forces i.e. the T=3/2 channel
is more sensitive to NNN forces in the tail region.
The import of this observation is that
it is the three-body breakup channels  which 
might show more evidence of 3BF since they
carry all of the T=3/2 strength. In fact analyzing our theoretical results
and the
experimental $nd$ and $ppn$ data from Ref. \cite{Fet}
leads us to conclude that in this energy
region the T=1/2 content of the $ppn$ channel is about 1/3.
Our findings
are in line with proposals \cite{ocon} which suggest
particular kinematical set-ups in the $nnp$ or $npp$ channels to be
sensitive to the presence of 3BF.

Figs.4(a) and 4(b) compare to available data \cite{Fet,FAUL81}
the total cross sections for $^3$H
and $^3$He as computed with two of our more complete
potential models, AV14+UrbVIII and Bonn-A + TM.
The $^3$He calculations include the full Coulomb interaction 
between the protons.  Our calculations agree rather closely with the
available data. We note by reference to Fig. 1 that the semi-realistic
models do not yield the same level of agreement with the data.

We have computed the total photo-absorption cross sections
of the trinucleons by using realistic NN and NNN forces.
Final state interactions have been fully included through the use
of the Lorentz integral transform method.
The results show clear sensitivity to the underlying nuclear
dynamics.  In particular the tail region of the T=3/2 channel
appears to be more sensitive to details of the NNN forces than to 
NN forces.  Therefore further theoretical and experimental
effort should be devoted to this promising kinematical region.
On the theoretical side several corrections, expected to be
minor, will be investigated in the future.  These are the
effects of higher multipoles, retardation in the one-body
operators and inclusion of explicit two-body currents
beyond the Siegert theorem.  In addition the calculation will
be extended to the new charge dependent potentials such as
the AV18 \cite{av18} and Nijmegen \cite{nijm} potentials
together with corresponding NNN models.
The rather large effect of the NNN force in the tail region
could imply a much larger effect in some selected kinematics
of the exclusive three-body breakup reaction.
In this regard it should be pointed out that a modification
of the Lorentz integral transform method used here
\cite{efros85,lapwl} will
allow the calculation of exclusive reaction cross sections
as well.

\begin{table}
\caption{ $^3$H Ground state properties for  AV14, Bonn-A,
AV14+UrbVIII, AV14+TM, and Bonn-A+TM potential models.}

\begin{tabular}{cccccc}
   Potential &E$_B$ (MeV)  &$\sqrt{<r_p^2>}$ (fm)  & P$_D$ & 
$\sigma_{-1}^{int}$ (mb) &$\sigma_{-1}^{gs}$ (mb) \\
\hline

AV14 & 7.69 & 1.66 & 8.94 & 2.636 & 2.645 \\
Bonn-A& 8.15 & 1.61 & 6.88 & 2.481 & 2.486 \\
AV14+UrbVIII & 8.49 & 1.60 & 9.67 & 2.442 & 2.453\\
AV14+TM      & 8.48 & 1.60 & 9.67 & 2.448 & 2.459 \\
Bonn-A +TM   & 8.47 & 1.59 & 7.25 & 2.411 & 2.415\\
\end{tabular}
\end{table}

\vfill\eject

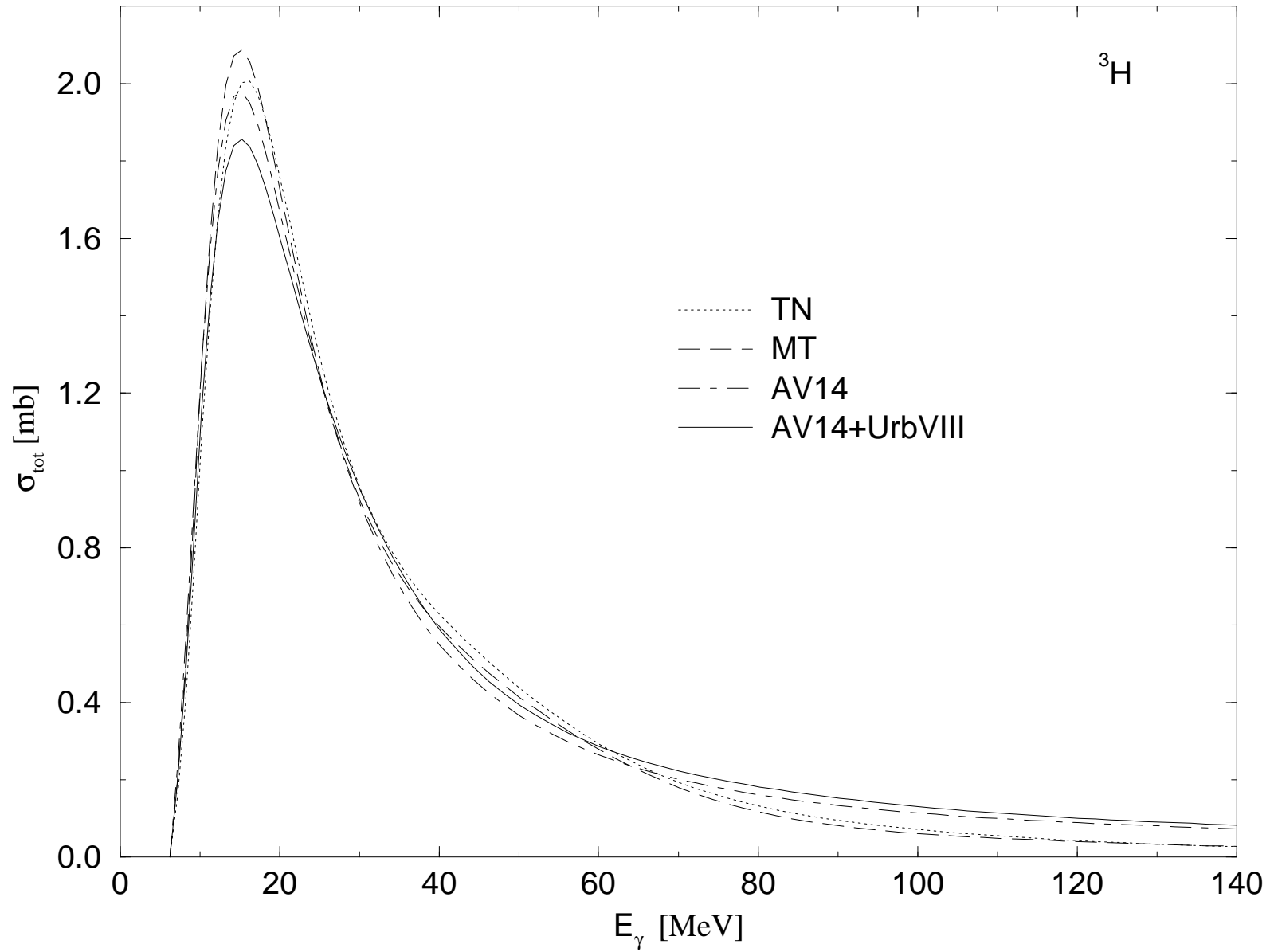
\begin{figure}
\caption{Total $^3$H photoabsorption cross sections predicted by
the TN (dotted curve), MT (dashed curve), AV14 (dash-dot curve) and
the AV14 + UrbVIII (solid curve) potential models.}
\end{figure}

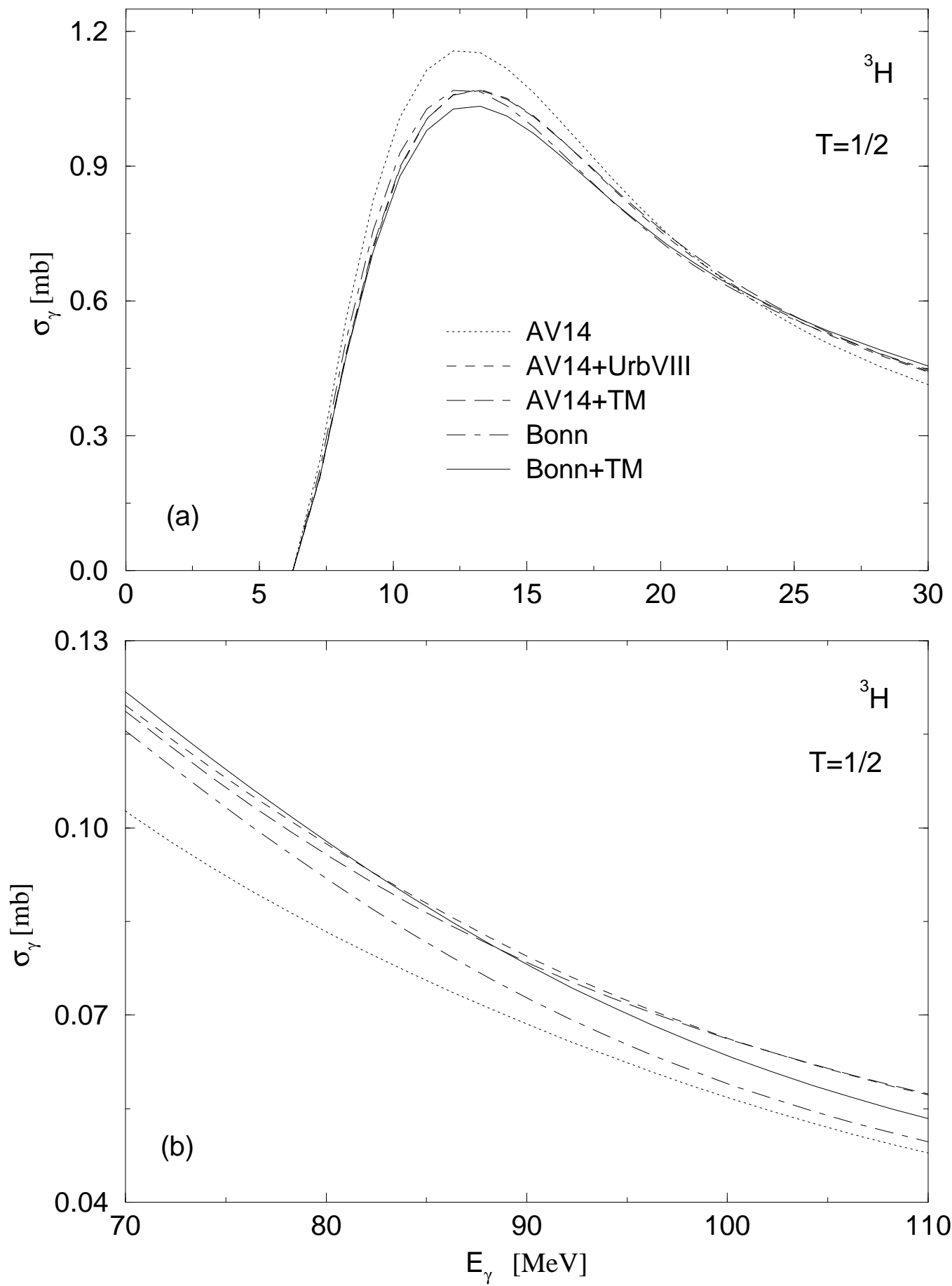
\begin{figure}
\caption{T=1/2 part of the $^3$H total photoabsorption cross sections
in the peak region
(a) and tail region (b).  The corresponding potentials are AV14 (dotted),
AV14 + UrbVIII (short dash), AV14 + TM (long dash), r-space Bonn-A 
(dash-dot), r-space Bonn-A + TM (solid).}
\end{figure}

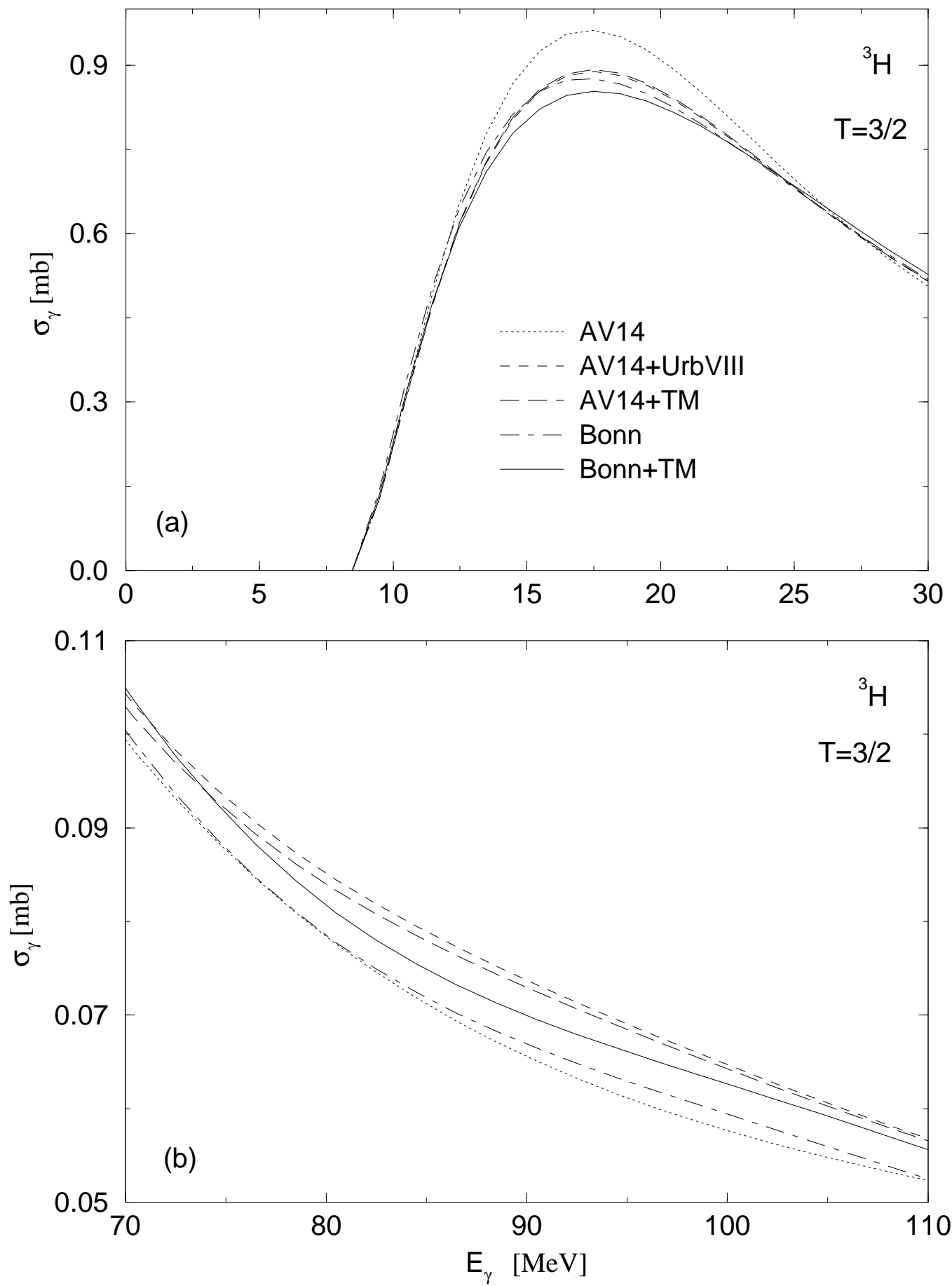
\begin{figure}
\caption{As in Fig. 2 but for T=3/2.}
\end{figure}

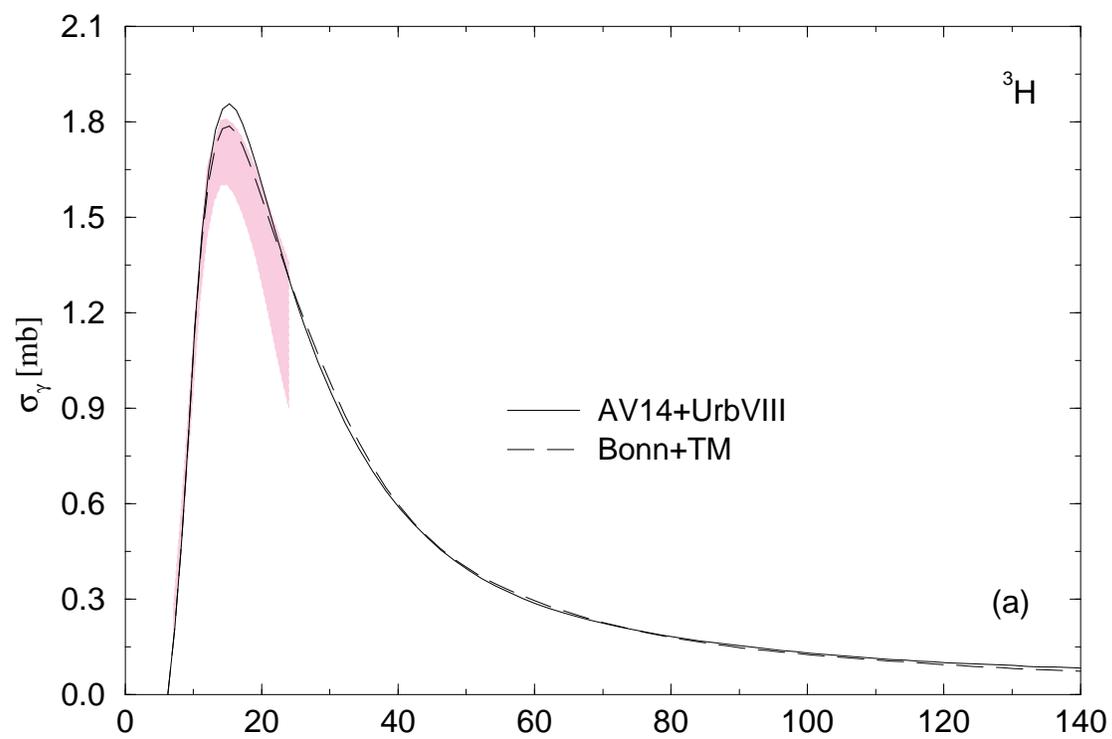
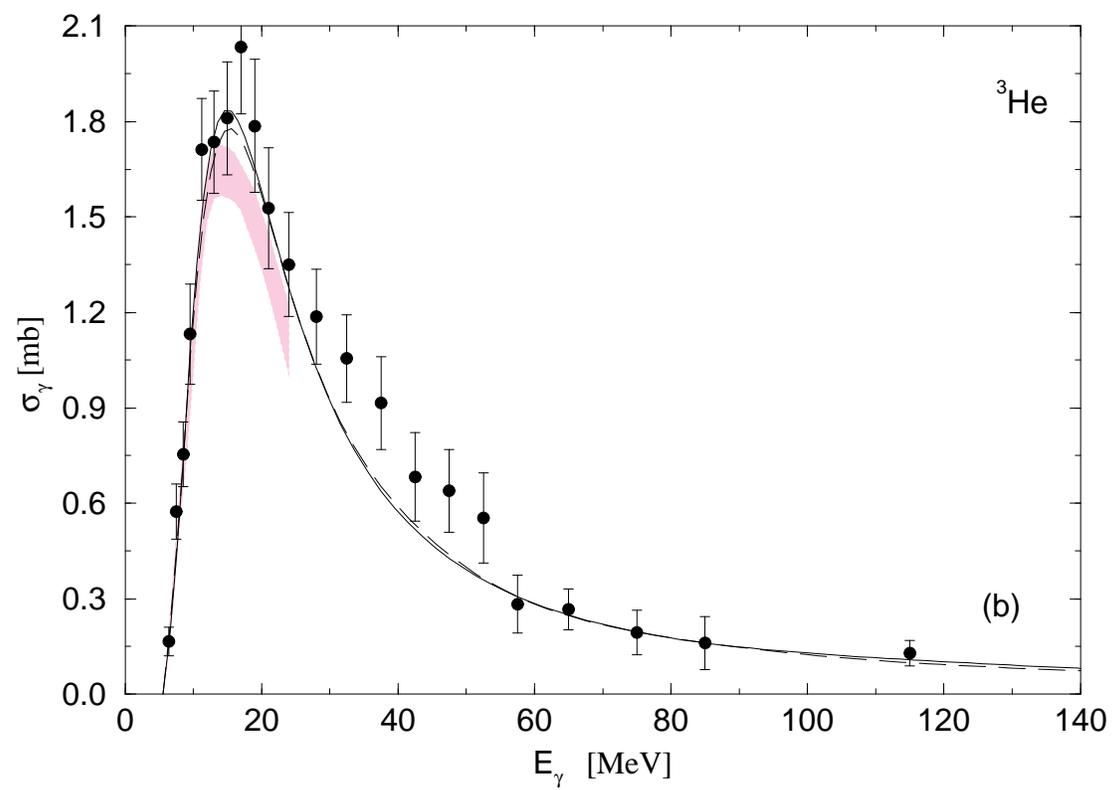
\begin{figure}
\caption{Total photoabsorption cross sections for $^3$H (a) and
$^3$He (b) computed from the AV14 + UrbVIII (solid) and
r-space Bonn-A + TM (dashed) models.
The shaded area represents the data from [19] while
the black dots are the data of [18]. }
\end{figure}

\section*{Acknowledgment}

The work of WL, GO, and ELT is partially supported by a NATO Collaborative
Research Grant. 
All authors acknowledge support from the Italian Ministery of Research (MURST).
In addition the work of ELT is
supported by the INFN and the National Science and Engineering Research Council
of Canada.

\end{document}